\documentclass[fleqn,10pt]{wlscirep}
\usepackage[utf8]{inputenc}
\usepackage[T1]{fontenc}
\usepackage{graphicx}
\usepackage{amsmath}
\usepackage{amssymb}
\usepackage{color}

\begin{document}
\title{
Inverse Renormalization Group based on Image Super-Resolution 
using Deep Convolutional Networks
} 

\author[1,2,*]{Kenta Shiina}
\author[1]{Hiroyuki Mori}
\author[3]{Yusuke Tomita}
\author[2,4,5,6]{Hwee Kuan Lee}
\author[1,+]{Yutaka Okabe}
\affil[1]{
Department of Physics, Tokyo Metropolitan University, 
Hachioji, Tokyo 192-0397, Japan}
\affil[2]{
Bioinformatics Institute, 
Agency for Science, Technology and Research (A*STAR),
30 Biopolis Street, \#07-01 Matrix, 
138671, Singapore}
\affil[3]{
College of Engineering, Shibaura Institute of Technology, 
Saitama 330-8570, Japan}
\affil[4]{
School of Computing, National University of Singapore, 
13 Computing Drive, 
117417, Singapore}
\affil[5]{
Singapore Eye Research Institute (SERI), 
11 Third Hospital Ave, 
168751, Singapore}
\affil[6]{
Image and Pervasive Access Laboratory (IPAL), 
1 Fusionopolis Way, \#21-01 Connexis (South Tower), 
138632, Singapore}

\affil[*]{16879316kenta@gmail.com}
\affil[+]{okabe@phys.se.tmu.ac.jp}

\def\l{\langle}
\def\r{\rangle}

\date{\today}

\begin{abstract}
The inverse renormalization group is studied based on the image 
super-resolution using the deep convolutional neural networks. 
We consider the improved correlation configuration instead of 
spin configuration for the spin models, such as the two-dimensional 
Ising and three-state Potts models. 
We propose a block-cluster transformation as an alternative to 
the block-spin transformation in dealing with the improved estimators. 
In the framework of the dual Monte Carlo algorithm, 
the block-cluster transformation is regarded as a transformation in the 
graph degrees of freedom, whereas the block-spin transformation 
is that in the spin degrees of freedom. 
We demonstrate that the renormalized improved correlation configuration 
successfully reproduces the original configuration at all the temperatures 
by the super-resolution scheme. Using the rule of enlargement, 
we repeatedly make inverse renormalization procedure 
to generate larger correlation configurations.  
To connect thermodynamics, an approximate temperature rescaling is discussed. 
The enlarged systems generated using the super-resolution 
satisfy the finite-size scaling.
\end{abstract}

\maketitle

\section*{Introduction}

Wilson revealed that the renormalization group (RG) is a key concept 
in understanding critical phenomena of phase transitions 
\cite{Wilson1,Wilson2}.
Kadanoff's block-spin transformation is an idea 
to realize renormalization in real space \cite{Kadanoff}.
The combination of the RG with Monte Carlo 
simulation has been successfully used as the Monte Carlo 
RG \cite{Ma,Swendsen,Pawley,Baillie}.
The inverse operation to generate a large system, which is 
called an inverse RG, was proposed by 
Ron, Swendsen, and Brandt \cite{Ron}. This approach is 
free of critical slowing down for large systems. 

Recent developments of machine-learning-based techniques 
have been applied to fundamental research, such as 
statistical physics \cite{Carleo}. 
A technique of supervised learning for image classification 
was used by Carrasquilla and Melko \cite{Carrasquilla} 
to propose a paradigm that is complementary 
to the conventional approach of studying interacting spin systems. 
By using large datasets of spin configurations, 
they classified and identified a high-temperature paramagnetic phase 
and a low-temperature ferromagnetic phase of 
the two-dimensional (2D) Ising model. 
Shiina {\it et al.} \cite{Shiina} extended and generalized this idea 
so as to treat various spin models including the multi-component 
systems and the systems with a vector order parameter. 
The configuration of a long-range spatial correlation 
was considered instead of the spin configuration itself. 
Not only the second-order and the first-order transitions 
but also the Berezinskii-Kosterlitz-Thouless (BKT) 
transition~\cite{Berezinskii1,Berezinskii2,kosterlitz,kosterlitz2} 
was studied. 

Tomita {\it et al.}~\cite{Tomita2020} have made further progress 
in this approach.  In the machine-learning study of 
the phase classification of spin models, the Fortuin-Kasteleyn (FK) 
\cite{KF,FK} representation-based improved estimators 
\cite{Wolff90,Evertz93} of the correlation configuration 
were employed as an alternative to the ordinary correlation configuration. 
This method of improved estimators was applied not only 
to the classical spin models but also to the quantum Monte Carlo 
simulation using the loop algorithm. 
They analyzed the BKT transition of the spin-1/2 quantum XY model 
on the square lattice. 

Another application of machine-learning study for image 
processing is super-resolution (SR), 
which is a class of techniques that enhance 
the resolution of an imaging system. 
Efthymiou {\it et al.}~\cite{Efthymiou} proposed a method 
to increase the size of lattice spin configuration using SR, 
deep convolutional neural networks \cite{Dong}. 
This study is related to the inverse RG 
approach~\cite{Ron}.  At high temperatures, however, there is 
a problem that the noise is largely random and difficult to learn. 
Because a significant reduction of variance is obtained 
for improved estimators at high temperatures of a disordered phase~\cite{Holm}, 
the improved correlation configuration could reduce 
the difficulty of SR. 

There have been some other proposals for the combination of neural network 
and RG~\cite{Beny,Mehta,Iso,Koch,Li}.  The work 
by Efthymiou {\it et al.}~\cite{Efthymiou} is an exceptional one 
to investigate an inverse RG. 

In this paper, we study the inverse RG of spin models 
based on the SR.  We consider the improved estimator 
of the correlation configuration instead of 
the spin configuration. 
As for the renormalization process, we propose 
a block-cluster transformation as an alternative to 
a block-spin transformation.  Then, we can set up 
an inverse RG procedure using the SR technique. 
The resolution of the enhanced configuration 
at high temperatures is much improved compared 
to the SR using the spin configuration.
We make inverse renormalization procedure repeatedly 
to generate larger correlation configurations.  
Introducing an approximate temperature rescaling, 
we show the finite-size scaling (FSS)~\cite{Fisher71,Fisher72,Binder} 
for the enlarged systems. 
For the spin models, we treat the 2D Ising model 
and the 2D three-state Potts model.

\section*{Results}

\subsection*{Monte Carlo renormalization group}

\vspace*{2mm}
\subsubsection*{Block-spin transformation}

We start with the RG process. To realize the RG in real space Monte Carlo 
simulation, Kadanoff's block-spin transformation \cite{Kadanoff} 
is conventionally used.  A majority rule is employed to determine 
a block spin from $2 \times 2$ spins in a block, for the square-lattice 
Ising model, for example. 

We do not consider the transformation of the Hamiltonian. 
Instead, we study the transformation property of the correlation. 
We calculate the correlation with a distance of $L/2$ 
as in the machine-learning study of the phase classification of spin models 
\cite{Shiina,Tomita2020}, where $L$ is a linear system size. 
This type of correlation function was used 
along with the generalized scheme for the probability-changing 
cluster algorithm~\cite{tomita2002b}. 
For actual calculation, we treat the average value 
of the $x$-direction and the $y$-direction for the site-dependent 
correlation, that is, 
\begin{equation}
  g_i(L/2) = (g[s_{x_i,y_i},s_{x_i+L/2,y_i}]+g[s_{x_i,y_i},s_{x_i,y_i+L/2}])/2,
\label{g_i}
\end{equation}
where $g[s, s']$ denotes a spin-spin correlation 
between a spin pair $s$ and $s'$.

\begin{figure}[t]
\begin{center}
\flushleft{(a) \hspace{7.4cm} (b)}

\includegraphics[width=7.4cm]{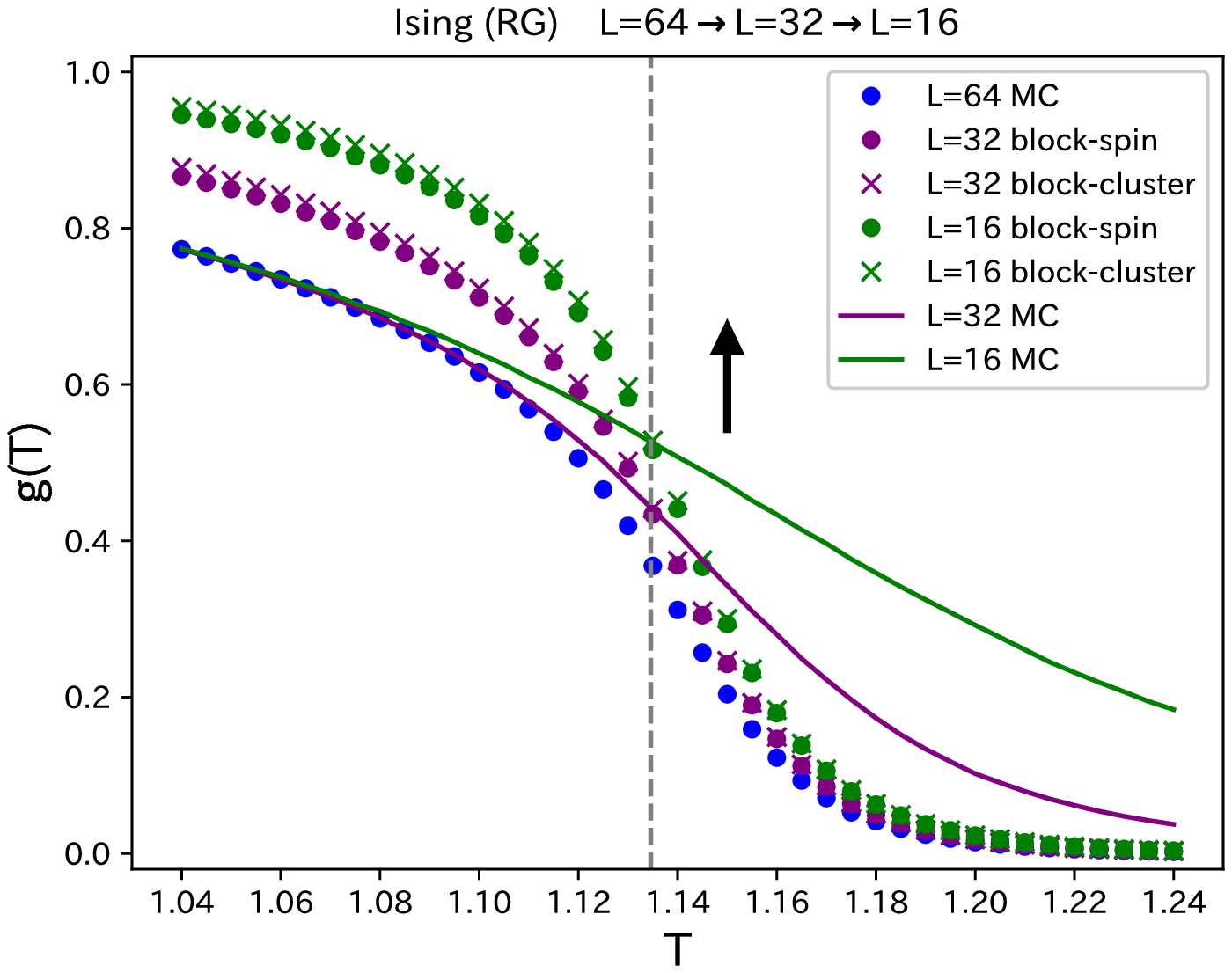}
\hspace{0.4cm}
\includegraphics[width=7.4cm]{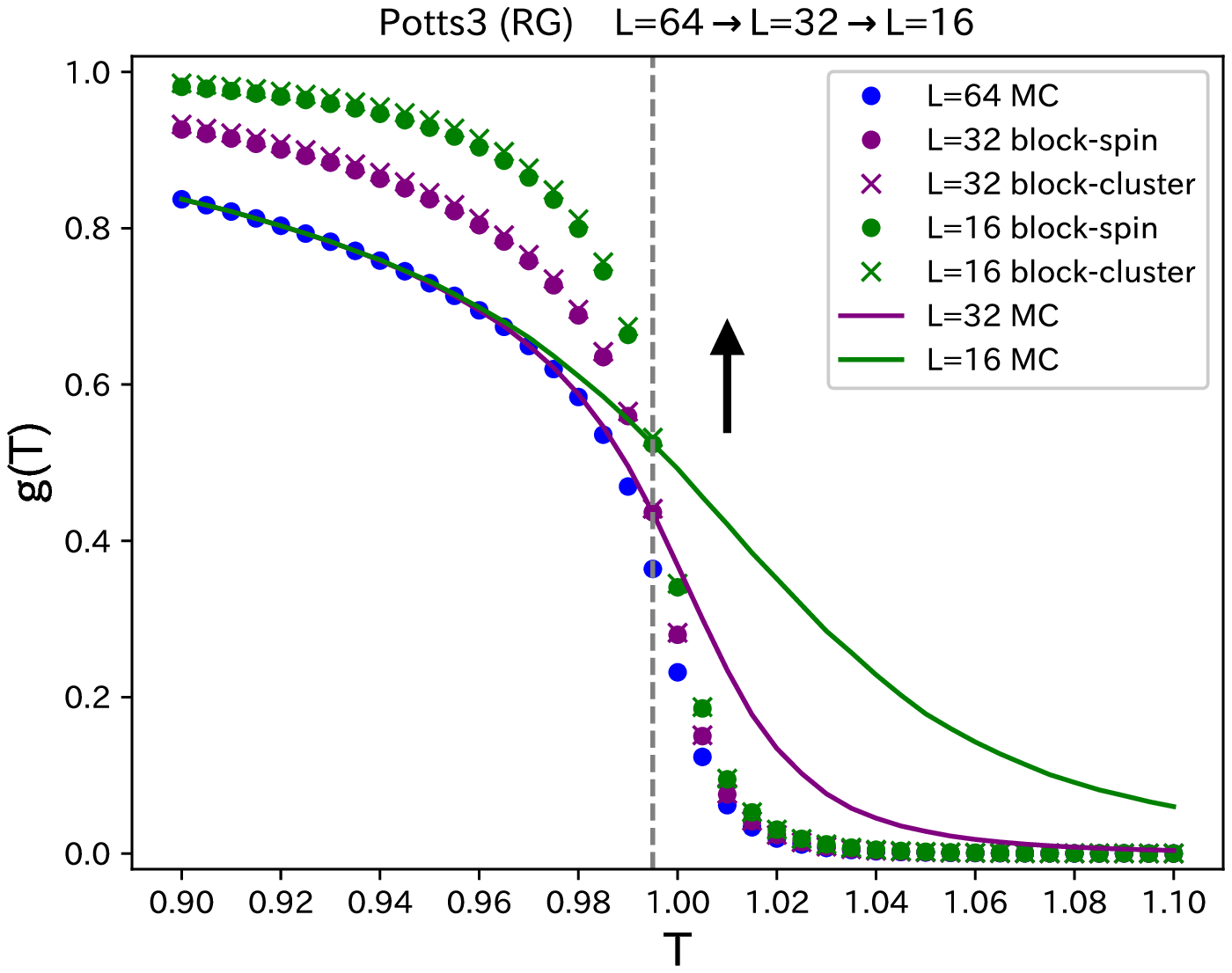}
\caption{
Renormalization-group procedures for (a) 2D Ising model 
and (b) 2D three-state Potts model. 
The temperature dependence of the correlation $g(T)$ 
for the $64 \times 64$ systems are plotted. 
The results of block-spin 
transformation (circles) and those of block-cluster 
transformation (crosses) are compared. 
The directions of renormalization are shown by arrows 
for convenience. 
We also plot the results of $32 \times 32$ and 
$16 \times 16$ systems by solid curves. 
The exact $T_c$'s are shown by dashed line.
}
\label{fig:RG}
\end{center}
\end{figure}

We performed the Monte Carlo simulation of the Ising model 
(2-state Potts model) on the $64 \times 64$ square lattice 
using the Swendsen-Wang cluster update~\cite{sw87}, 
and made block-spin transformations repeatedly. 
When a block-spin transformation is made one time, 
the linear system size becomes a half of the original size.
We measured the space-averaged value of $g_i(L/2)$ 
for the original Ising spins and also for the block spins. 
We plot the temperature ($T$) dependence of the total 
correlation, 
\begin{equation}
  g(T) = \l \frac{1}{N} \sum_{i=1}^N g_i(L/2) \r,
\end{equation}
by circles in Fig.~\ref{fig:RG}(a), where $N (=L \times L)$ is the 
system size, and the angular brackets denote 
the Monte Carlo average.  The temperature is measured 
in units of the interaction $J$ (in terms of the Potts model). 
We also plot the temperature dependence $g(T)$ of 
the Ising model with $L=32$ and $L=16$ by solid curves. 
We observe that the block-spin correlation $g(T)$ of $L=32$ 
produced from $L=64$ system and the correlation $g(T)$ 
of the true $L=32$ system cross at the exact 
$T_c = 1/\ln(1+\sqrt{2}) = 1.1346 \cdots$, which is shown 
by vertical dashed line. 
The two-time block-spin correlation $g(T)$ of $L=16$ 
and the correlation of the true $L=16$ system cross at the 
exact $T_c$. It is noteworthy that there are some corrections to scaling. 

As another example, we treated the three-state Potts model 
on the square lattice.  For the block-spin transformation, 
we use a majority rule. 
We plot the temperature dependence $g(T)$ of the three-state Potts model 
on the square lattice with $L=64$ in Fig.~\ref{fig:RG}(b). 
The block-spin correlations are compared with the correlations 
of the true $L=32$ and $L=16$ systems as in the case of 
the Ising model, Fig.~\ref{fig:RG}(a). We again observe the crossing 
at the exact $T_c$, which is $T_c = 1/\ln(1+\sqrt{3}) = 0.9950 \cdots$ 
for the three-state Potts model.

\subsubsection*{Block-cluster transformation}

In the development of the cluster update of the Monte Carlo 
simulation, the so-called improved estimators~\cite{Wolff90,Evertz93} 
were proposed for the measurement of the correlation. 
In calculating spin correlations, only the spin pair  
belonging to the same FK cluster should be considered.  
In the improved estimator for the cluster representation of 
the $q$-state Potts model (including the Ising model), 
the correlation becomes 1 for the spin pair belonging to 
the same FK cluster, whereas it becomes 0 for the spins 
of different clusters. 
In the framework of the dual Monte Carlo 
algorithm~\cite{Kandel,Kawashima95,Kawashima04}, 
the Markov process in the cluster update alternates 
between the original spin configurations (spin) and the space of 
the configurations of auxiliary variables (graph). 
Then, an improved estimator is considered to be an estimator 
defined in terms of the graph degrees of freedom rather 
than the original spin degrees of freedom. 

In the proposal of using SR by Efthymiou {\it et al.}~\cite{Efthymiou}, 
there is a problem that the noise is largely random and difficult 
to learn at high temperatures. 
We will discuss in the present paper that the improved correlation 
configuration solves this difficulty. 
However, we cannot directly apply the improved estimator 
for the block-spin transformation; we cannot specify 
the FK cluster to which a selected block spin belongs. 
Thus, we here propose another renormalization procedure, 
a block-cluster transformation. 
The detailed procedure of a block-cluster transformation
is described in the section of Methods.

We performed the block-cluster transformation for the 2D Ising 
model and the three-state Potts model. 
The plots of the temperature dependence of the correlation $g(T)$ 
are given by crosses in Figs.~\ref{fig:RG}(a) and \ref{fig:RG}(b). 
We may compare the results of the block-spin transformation (circles) 
and those of the block-cluster transformation (crosses). 
We observe that the renormalized values of $g(T)$ are almost the same 
for both the Ising model and the three-state Potts model. 
At high temperatures fluctuations of the block-cluster 
transformation are smaller than those of the block-spin 
transformation because the improved estimators are used 
in the block-cluster transformation.
There are very small deviations at low temperatures, 
which depend on the renormalization scheme. 
At $T_c$ of the Ising model, the block-cluster value of $g(T_c)$ 
for $L=32$ (purple cross) is 1.6\% larger than the block-spin value 
(purple circle), whereas the true value of $L=32$ (purple curve) 
is between, and close to the value of the block-cluster transformation. 
From the viewpoint of transformation property of renormalization, 
the block-cluster transformation could be better.

\section*{Inverse renormalization group based on super-resolution}

\begin{figure}[t]
\begin{center}
\flushleft{(a) \hspace{7.4cm} (b)}

\includegraphics[width=7.4cm]{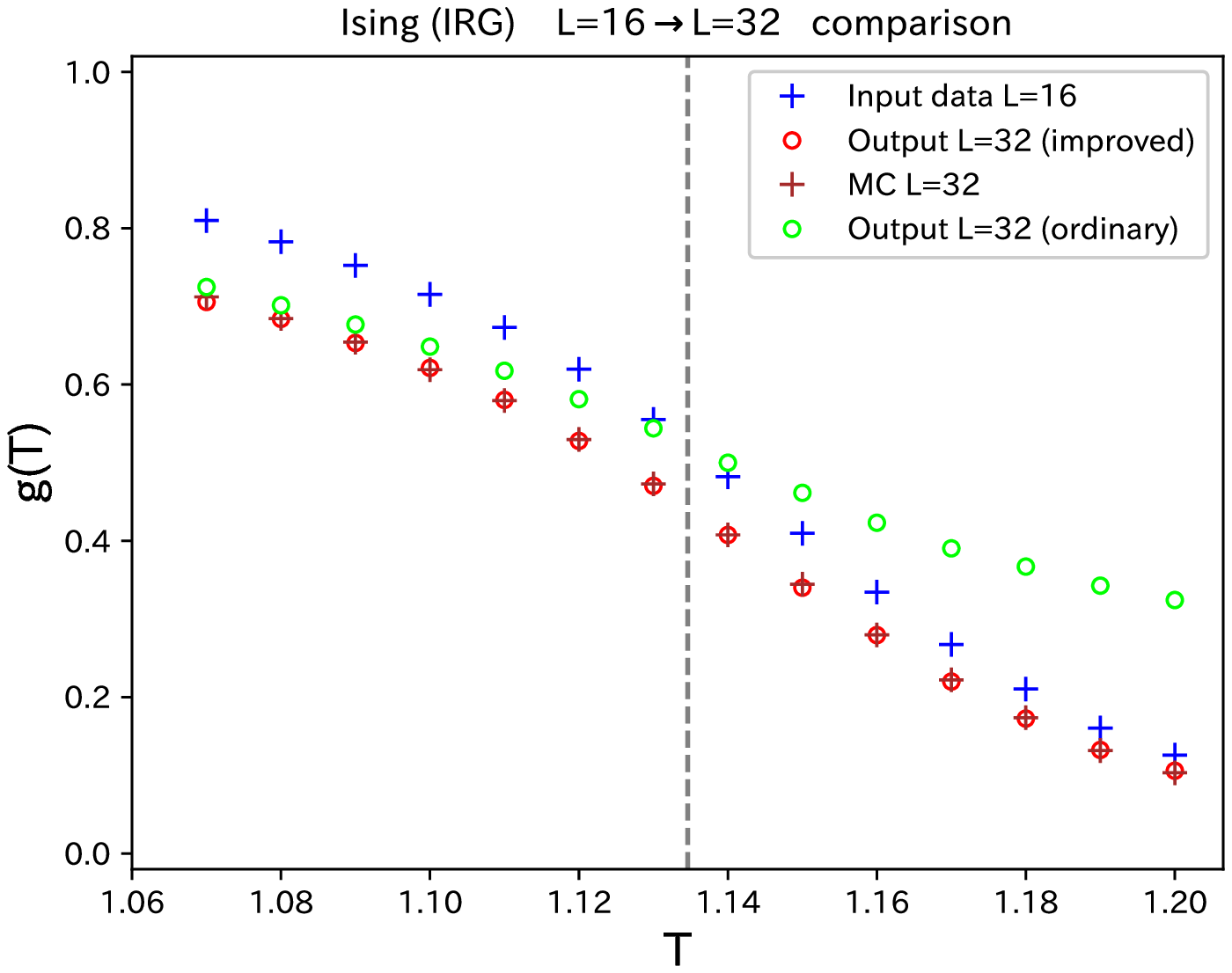}
\hspace{0.4cm}
\includegraphics[width=7.4cm]{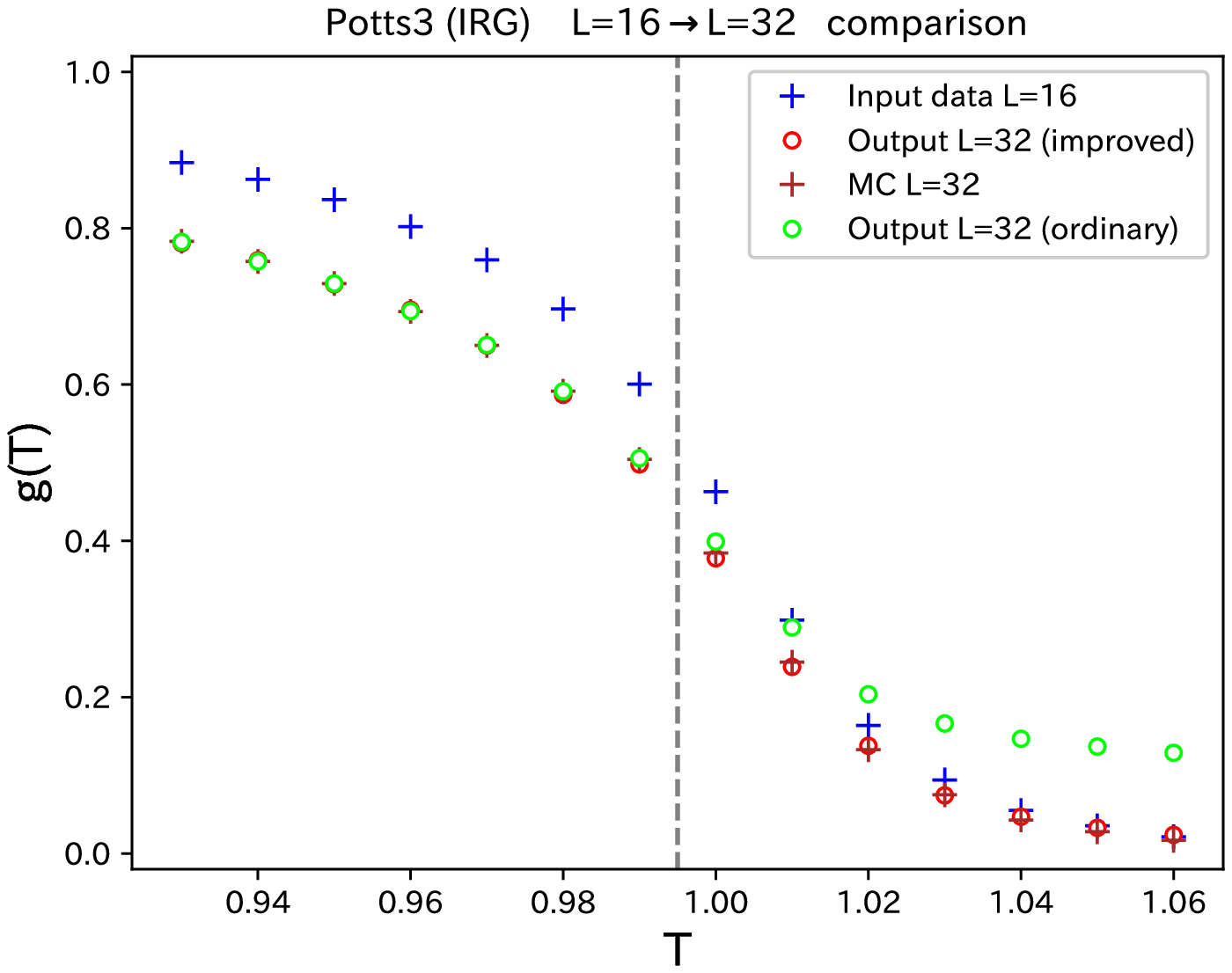}
\caption{
Image super-resolution for (a) 2D Ising model 
and (b) 2D three-state Potts model. Starting with the truncated correlation configuration ($L=16$) 
(blue pluses), we compare $g(T)$ of the original configuration 
of test data ($L=32$) (brown pluses) with the reproduced $g(T)$ 
(red circles).  For comparison, we also plot the results of 
SR for the ordinary correlation configuration produced 
from the spin configuration (lime circles). 
The exact $T_c$'s are shown by dashed line.
}
\label{fig:iRG_comparison}
\end{center}
\end{figure}

\subsubsection*{Extension of the method of Efthymiou $\boldsymbol{et \ al.}$}

\begin{figure}[t]
\begin{center}
\flushleft{(a) \hspace{7.4cm} (b)}

\includegraphics[width=7.4cm]{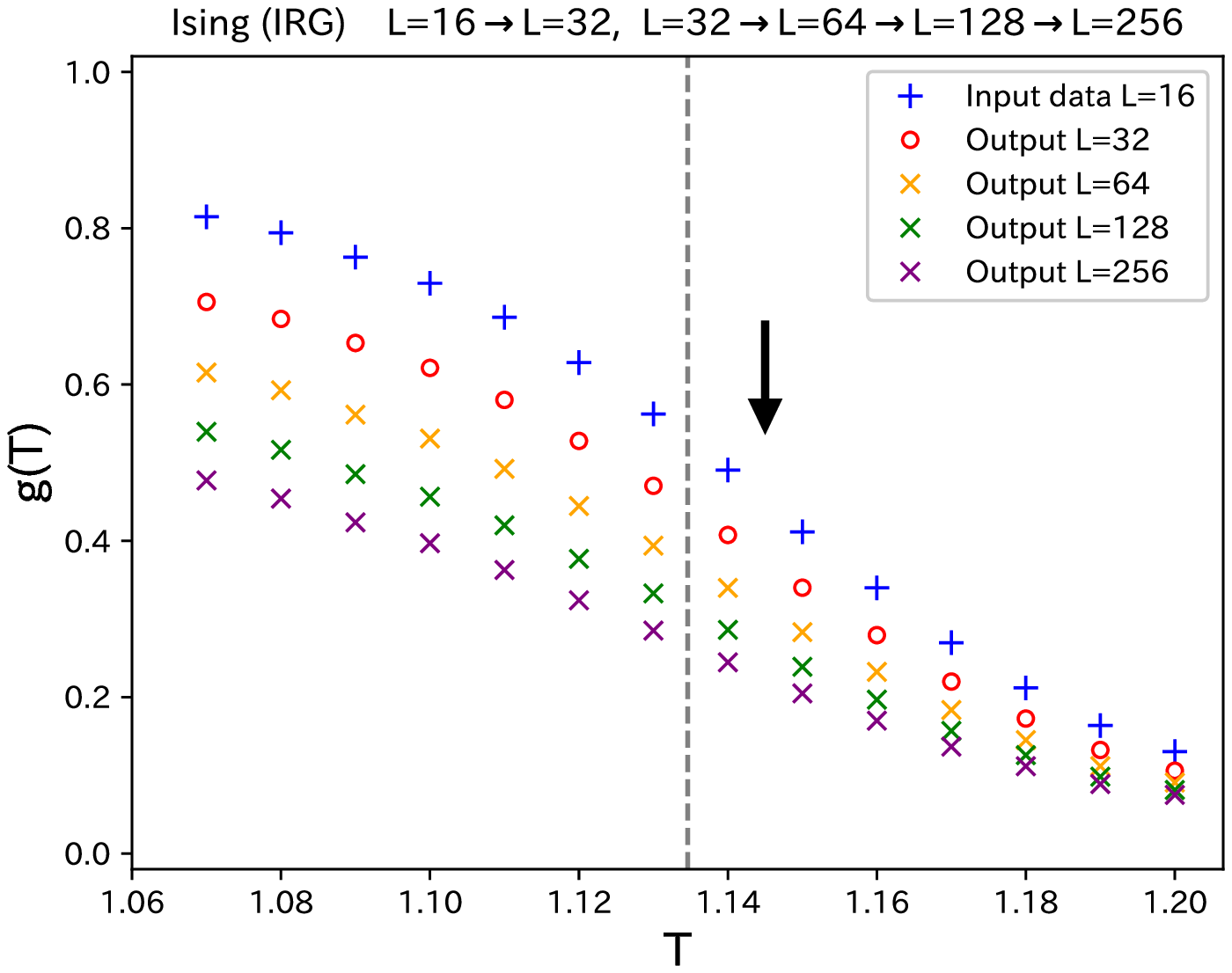}
\hspace{0.4cm}
\includegraphics[width=7.4cm]{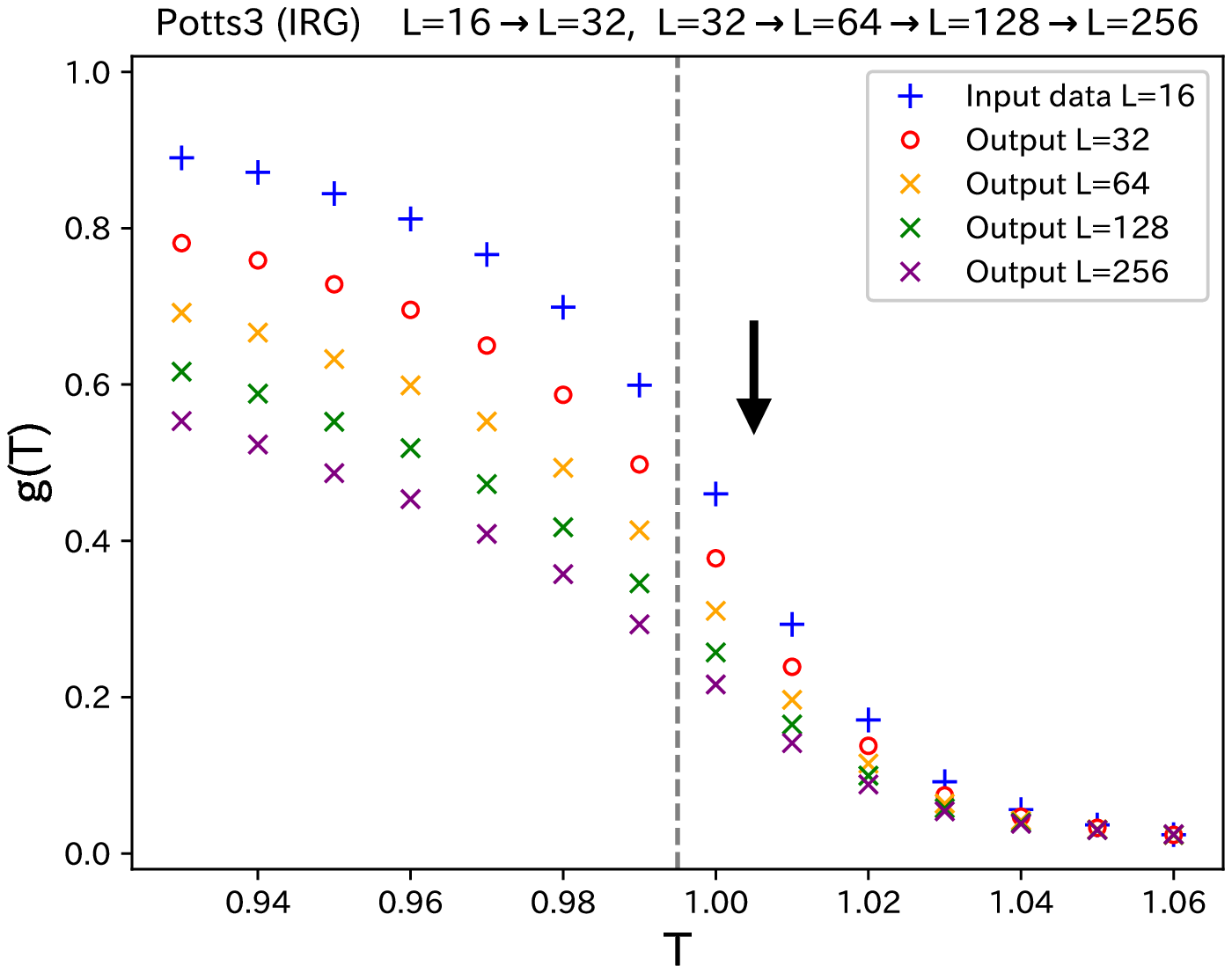}
\caption{
Iterative inverse renormalization-group procedures for (a) 2D Ising model 
and (b) 2D three-state Potts model. 
Starting with the truncated correlation configuration ($L=16$) 
(blue pluses), we obtain the reproduced $g(T)$ of $L=32$ (red circles).  
The enlarged $g(T)$'s are plotted by orange crosses 
($L=64$), green crosses ($L=128$), and purple crosses ($L=256$). 
The directions of inverse renormalization are shown by arrows 
for convenience, which are opposite to the renormalization 
shown in Fig.~\ref{fig:RG}. 
The exact $T_c$'s are shown by dashed line.
}
\label{fig:iRG}
\end{center}
\end{figure}

We perform the inverse operation of RG, 
extending the method of SR with a deep convolutional neural 
network (CNN) \cite{Dong} due to Efthymiou {\it et al.}~\cite{Efthymiou}. 
We use the improved estimator of the correlation configuration 
instead of the spin configuration. 
We give the detailed description of SR 
in the section of Methods. 
We emphasize that the present method can be applied to any $q$-state 
Potts model because we deal with the correlation configuration. 
In the case of the spin configuration, only the Ising model 
can be treated. 

We performed the procedure of SR for the 2D Ising model 
and three-state Potts model. 
Simulating $32 \times 32$ systems, 
we obtain sets of original improved correlation configuration 
($32 \times 32$) and the truncated improved correlation configuration 
($16 \times 16$) using block-cluster transformation. 
For training data we use 8000 sets of configurations for each temperature. 
In Fig.~\ref{fig:iRG_comparison}, we plot the sample average 
of $\sum_i \xi_i$ and $\sum_i g_i$; that is, $g(T)$ for $L=32$ 
(brown pluses) and $L=16$ (blue pluses). 
Here, $\xi_i$ and $g_i$ are original and truncated improved 
correlation configurations, respectively. 
Using the SR technique, the parameters ($\theta = (W,b)$) are 
tuned for each temperature. 
For test data we use other independent 6000 sets of original and 
truncated configurations. 
The improved correlation configuration of $L=32$ is reproduced 
from the $L=16$ truncated configuration 
using the optimized parameters $\theta$. 
We compare the original $\l \sum_i \xi_i \r$ (brown pluses) 
and the reproduced $\l \sum_i \xi_i' \r$ (red circles) 
in Fig.~\ref{fig:iRG_comparison}. 
We do not observe appreciable differences; it means that 
the reproduction is almost perfect for all the temperatures. 
In Fig.~\ref{fig:iRG_comparison}, for comparison, we also plot 
the results of SR for the ordinary correlation configuration produced from 
the spin configuration (lime circles). 
At high temperatures, there is a deviation from the 
true values of $L=32$ (brown pluses). 
To avoid this difficulty, an additional term was 
added in the regularization term in the loss function 
in Ref.~\citen{Efthymiou}. 
We do not need to add such an additional term for the improved estimator. 
The relative deviation of reproduction $\l \sum_i \xi_i \r$
$\to$ $\l \sum_i \xi_i' \r$ is smaller than 0.4\% 
at all the temperatures. 

Next consider the further increase of the system size. 
From the improved correlation configuration of $32 \times 32$ with 
tuned parameters $\theta$, we generate the SR configuration of 
$64 \times 64$.  
The $2L \times 2L$ output of the first SR is regarded as 
the input of a new network $2L \times 2L \to 4L \times 4L$. 
Following Efthymiou {\it et al.}~\cite{Efthymiou}, 
we assume that the parameters (the weight matrix $W$ and the bias 
vector $b$) are independent of the system size. 
In the same way, we generate $128 \times 128$ enlarged image, 
$256 \times 256$ enlarged image, repeatedly. 
In Fig.~\ref{fig:iRG}, we plot the iterative SR for both 2D Ising model 
and 2D three-state Potts model.  It is noteworthy that 
we make Monte Carlo simulations only for the system size 
of $32 \times 32$.  This SR procedure is a geometric procedure, 
and temperature has its own meaning only for $32 \times 32$ system. 

\begin{figure}[t]
\begin{center}
\flushleft{(a) \hspace{7.4cm} (b)}

\includegraphics[width=7.4cm]{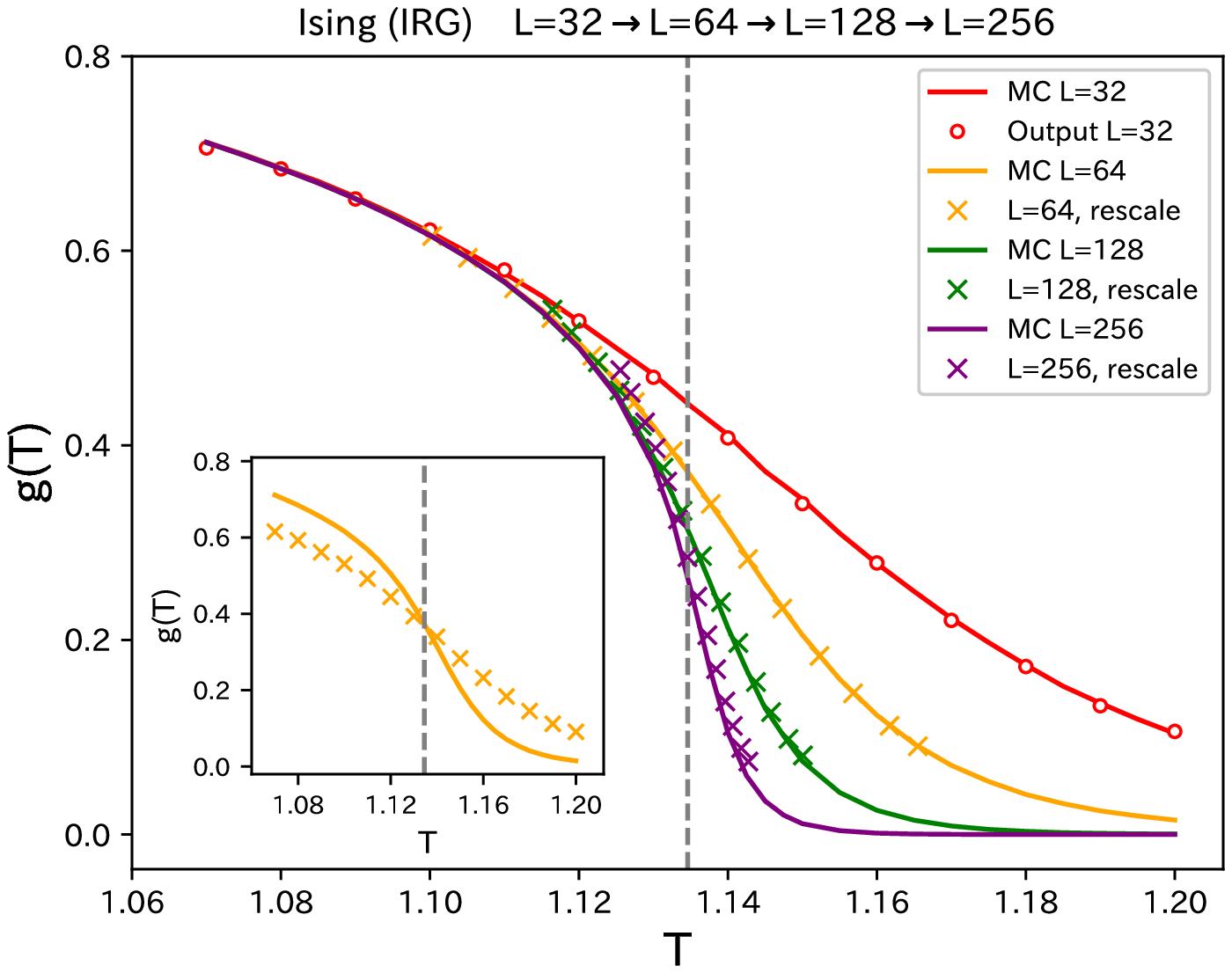}
\hspace{0.4cm}
\includegraphics[width=7.4cm]{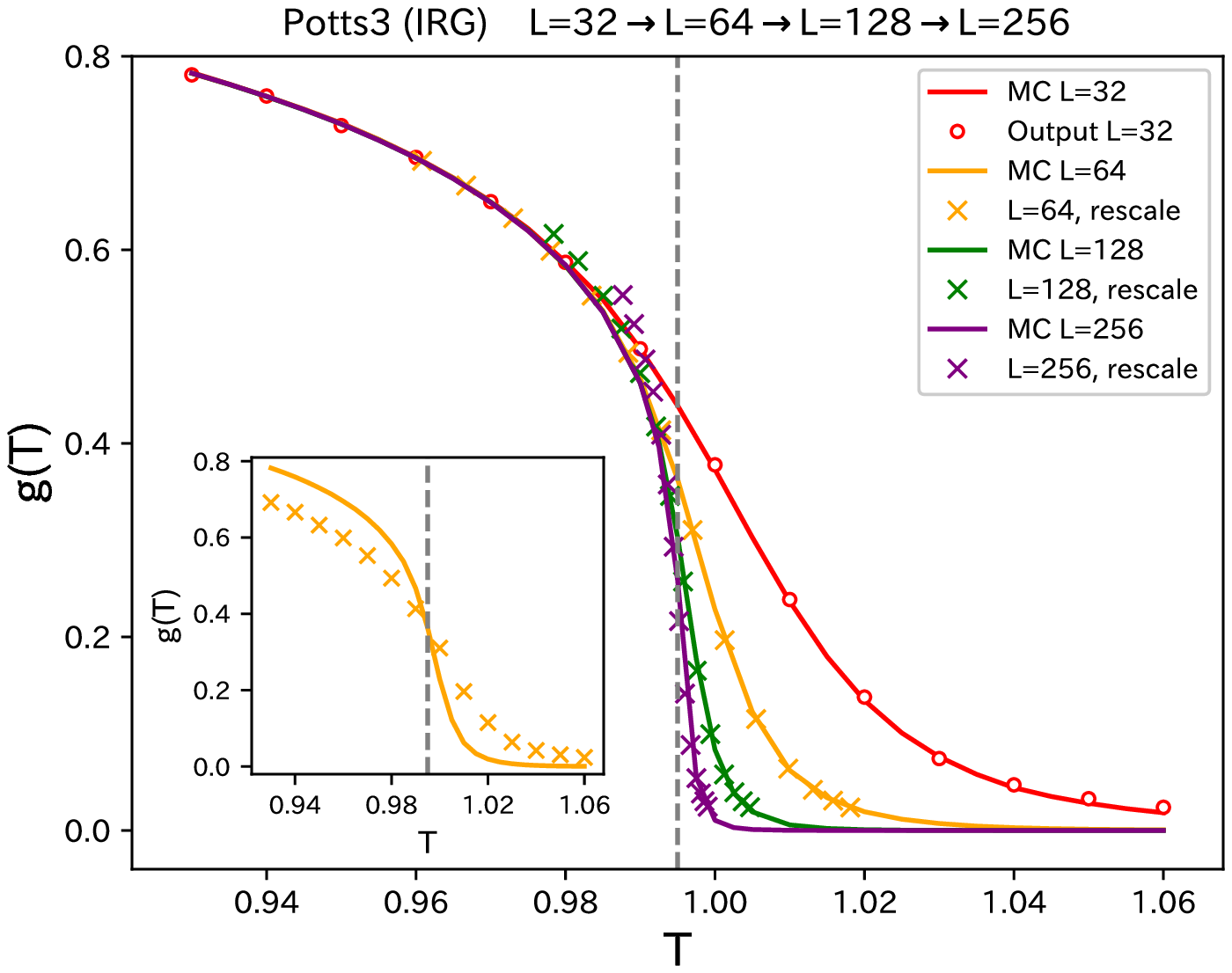}
\caption{
Approximate temperature rescaling for (a) 2D Ising model 
and (b) 2D three-state Potts model.
The transformation rule $\tilde T \to T$ is established by requiring 
the correlation $g(T)$ for enlarged system of $L=64$ (orange crosses) 
and corresponding $g(T)$ curves of true $L=64$ system (orange curve) 
to collapse, 
which is shown in the inset of the plots in the non-rescaled $T$. 
In the main figure, the $g(T)$'s of $L=64$ (orange crosses) are plotted 
as a function of the rescaled $T$. 
By using this rule, the $g(T)$'s of enlarged systems of $L=128$ and $L=256$ 
are rescaled (crosses), and compared with the true correlations (solid curves). 
The exact $T_c$'s are shown by dashed line.
}
\label{fig:rescaling}
\end{center}
\end{figure}

\subsubsection*{Temperature rescaling}

For the one-dimensional Ising model, self-similar transformation 
of the Hamiltonian using the decimation scheme is possible. 
However, the nearest-neighbor 2D Ising model will be 
mapped to the model with complex interactions, such as 
the next-nearest-neighbor interaction, four-spin interaction, etc. 
Thus, for the 2D models, the transformation of Hamiltonian 
is not self-similar after a block-spin RG transformation, 
and therefore temperature alone is not sufficient to describe
the coupling space of the RG configuration. 
We follow Efthymiou {\it et al.}~\cite{Efthymiou} to describe 
a method to approximate the rescaling of temperature numerically. 
We transform temperature such as $\tilde T = F(T)$ 
with the RG transformation. To avoid confusion 
with the activation function $f$, we here use $F$ 
instead of $f$. In opposite direction 
of inverse RG transformation, transformation 
such as $T = F^{-1}(\tilde T)$ is expected. 
To find the rescaling, 
we compare the correlation $g(T)$ calculated for enlarged system 
of $64 \times 64$ with the true $g(T)$ of $64 \times 64$ system. 
We find the transformation $\tilde T \to T$ by requiring the
corresponding $g(T)$ curves to collapse. 

Once the transformation $T = F^{-1}(\tilde T)$ from $32 \times 32$ 
to $64 \times 64$ systems is established, we use the same transformation 
to the inverse RG procedure of $64 \to 128$ and $128 \to 256$. 
In Fig.~\ref{fig:rescaling}, we show the temperature rescaling 
results. The temperatures of the output of SR for $64 \times 64$ 
(orange crosses) are rescaled such that they collapse 
with the true $g(T)$ obtained by the Monte Carlo simulation 
of $64 \times 64$ (orange curve), which is shown in the inset.  
The outputs of $128 \times 128$ and $256 \times 256$ are rescaled using 
the transformation $F^{-1}$ obtained by $32 \to 64$ transformation. 
We observe that the range of the rescaled temperature shrinks.
Monte Carlo results of $128 \times 128$ and $256 \times 256$ 
are also given in Fig.~\ref{fig:rescaling} for the sake of comparison. 
For high temperature sides, temperature rescaling works quite well. 
However, for low temperature side, deviation becomes appreciable 
when the temperature becomes away from $T_c$.  It comes 
from the saturation effects of $g(T)$, that is, it approaches 1 
as $T \to 0$.  This phenomena often appear in FSS
analysis of magnetization, for example.

For geometric inverse RG, only the Monte Carlo simulation 
of $32 \times 32$ system is enough. 
For temperature rescaling, Monte Carlo simulation of $64 \times 64$ 
should be added.  With these small sizes of simulation, 
we can obtain the information on larger system sizes.

\begin{figure}[t]
\begin{center}
\flushleft{(a) \hspace{7.4cm} (b)}

\includegraphics[width=7.4cm]{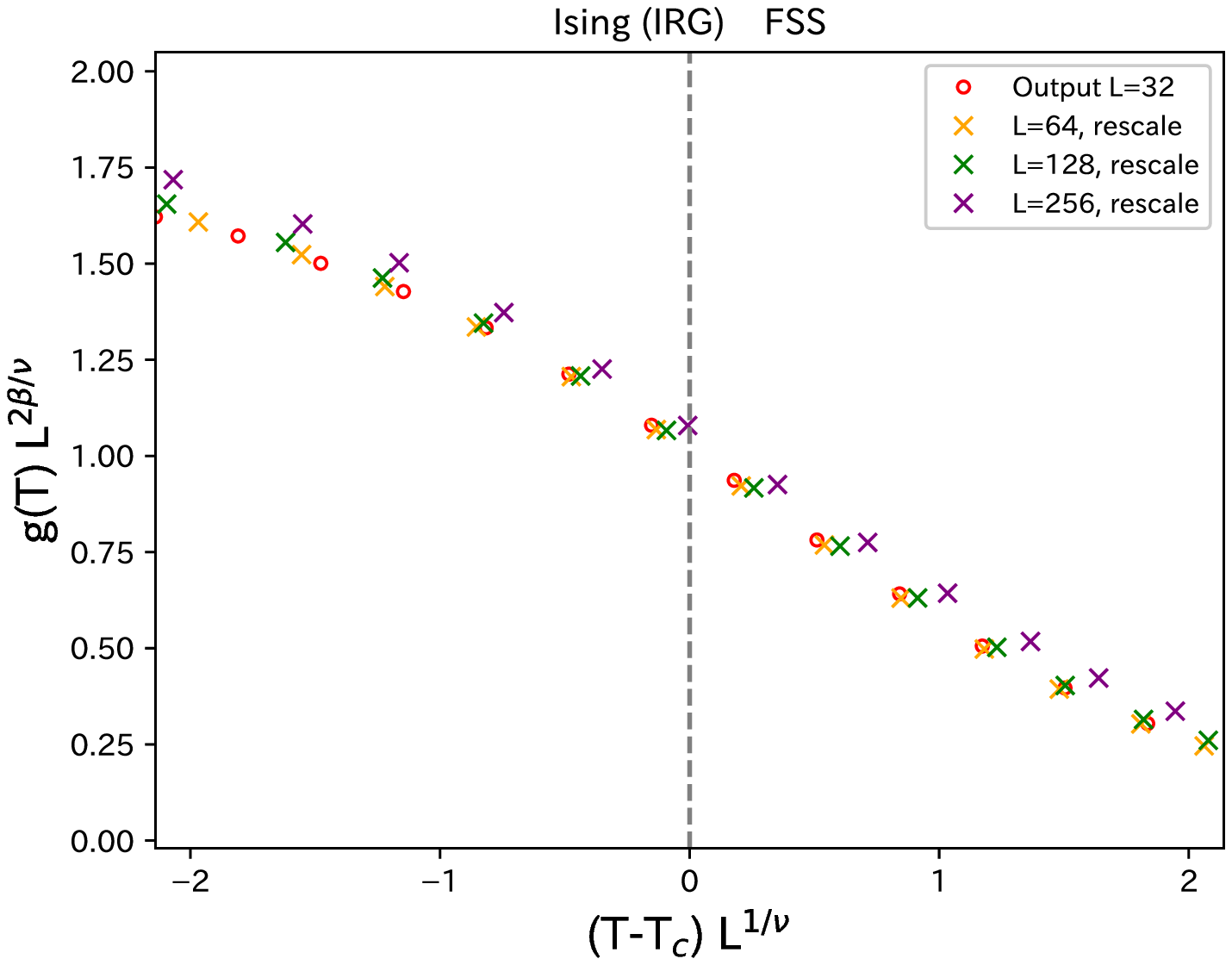}
\includegraphics[width=7.4cm]{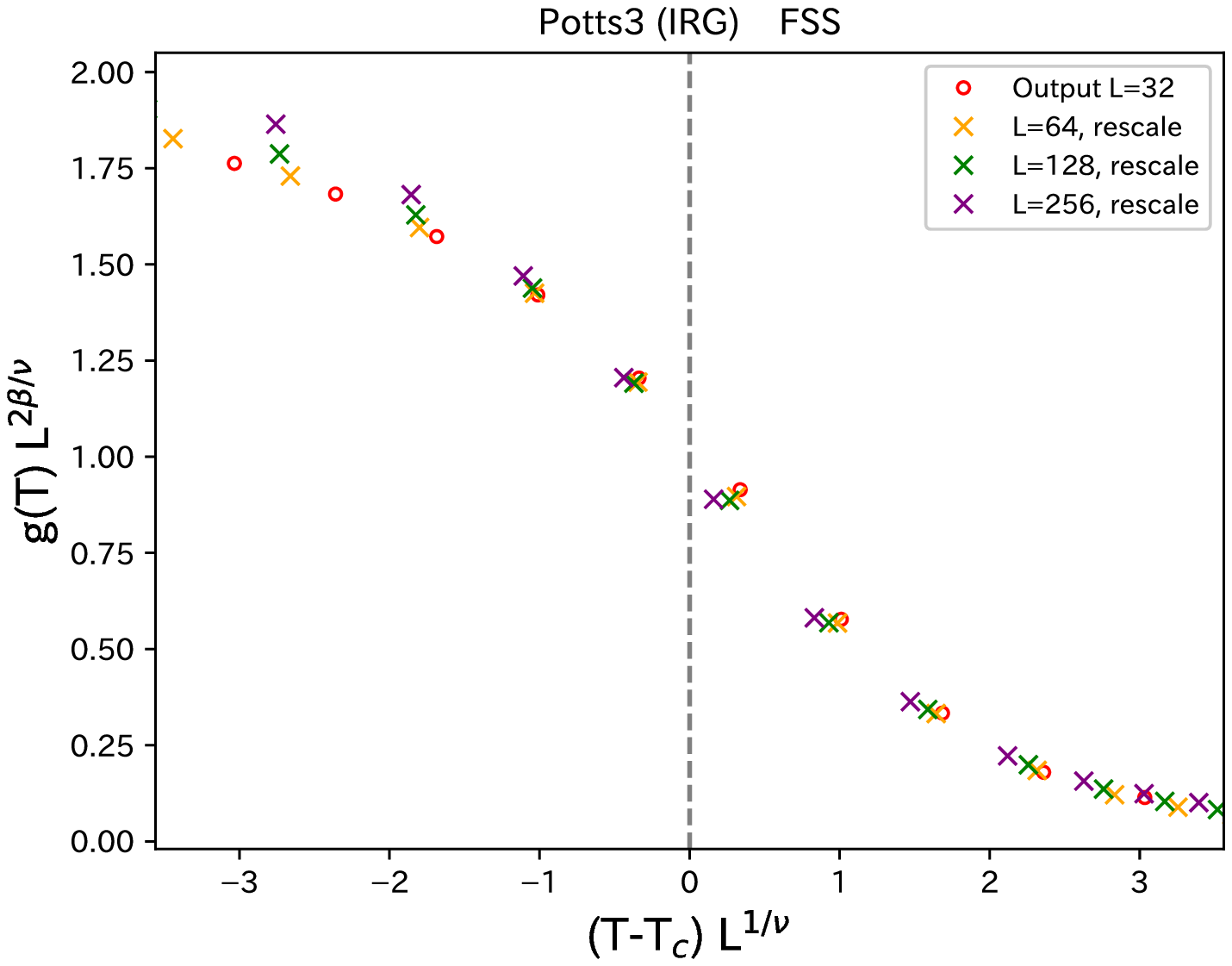}
\caption{
FSS for (a) 2D Ising model 
and (b) 2D three-state Potts model.
The temperature-rescaled data shown in Fig.~\ref{fig:rescaling} 
are plotted. 
As for $T_c$, we used the exact values because our estimated 
fixed points, $T^* = F^{-1}(T^*)$, are very close to 
the exact values, as shown in the inset of Fig.~\ref{fig:rescaling}. 
The critical exponents $2\beta/\nu$ and $1/\nu$ 
are chosen, such that 
data up to $L=128$ are collapsed. 
They are 0.240 (1/4) and 1.010 (1)
for the Ising model, and 0.255 (4/15) and 1.215 (6/5) 
for the three-state Potts model. In the parentheses, 
the exact exponents are given.
}
\label{fig:FSS}
\end{center}
\end{figure}

\subsubsection*{Finite-size scaling}

Now that we have obtained the temperature dependence $g(T)$ 
for various sizes from small sizes, we try a FSS 
analysis~\cite{Fisher71,Fisher72,Binder} to examine critical phenomena. 
The FSS function for the equation of state can be written as 
\begin{equation}
  g(T) L^{2\beta/\nu} = \tilde g(t L^{1/\nu}), 
\end{equation}
where $t=T-T_c$ with the critical temperature $T_c$; 
$\nu$ and $\beta$ are the correlation-length and magnetization 
critical exponents, respectively. 
In Fig.~\ref{fig:FSS}, we show the FSS plots of the Ising model 
and the three-state Potts model. That is, $g(T) L^{2\beta/\nu}$ 
is plotted as a function of $t L^{1/\nu}$. 
As for the critical temperature $T_c$, we use the exact values 
because in the temperature rescaling transformation shown in 
the inset of Fig.~\ref{fig:rescaling}, the fixed points, 
$T^* = F^{-1}(T^*)$, are very close to the exact values. 
For the critical exponents, $2\beta/\nu (= \eta)$ and $1/\nu$, 
best-fitted values for data collapsing up to $L=128$ were used. 
The chosen $2\beta/\nu$ and $1/\nu$ 
are 0.240 (1/4) and 1.010 (1) for the Ising model, 
and 0.255 (4/15) and 1.215 (6/5) 
for the three-state Potts model.  
In the parentheses, the exact exponents are given. 
Estimated critical exponents are $1 \sim 4 \%$ accuracy. 
We obtained good FSS, although the original system of small size 
of $L=32$ has larger corrections to FSS. 

\section*{Discussion}

We have successfully realized the inverse RG based on 
SR approach.  We have proposed the block-cluster transformation 
as an alternative to the block-spin transformation to use 
the improved estimators. 
We have made inverse renormalization procedure repeatedly 
to generate larger correlation configurations.  
We here make remarks on the advantage of the present method 
and the future directions of the research.

The advantage of using cluster representation of spin models 
is remarkable, and the improved estimator is quite useful. 
In doing so, we have introduced a block-cluster transformation, 
and we investigate the improved correlation configuration. 
The statistical advantage of improved estimator is well known~\cite{Holm}; 
at high temperatures above $T_c$, the errors for the spatial 
average of correlation is drastically reduced.  
In addition, this study elucidated 
the advantage in improved correlation configuration itself.

We make comments on the image as an object of image processing. 
Images usually have some smooth parts together with some edges. 
Abstract painting is not an object of image processing. 
A similar example is found in the case of text compression; 
we discuss the compression of text in natural language, 
while the compression of random sequences is impossible. 
The spin configurations have particular characteristics. 
There is a long-range correlation up to the correlation length. 
Moreover, there are some symmetries.  The ordered-state 
spin configurations at low temperatures have some smooth parts. 
However, high-temperature spin configurations are random. 
On the other hand, improved correlation configurations have 
cluster structures even at high temperatures. 
The choice of proper "image" is important in applying the technique of 
image processing to the spin configuration problems. 

The RG transformation is a coarse-grained procedure; thus, 
the truncated systems (block-spin or block-cluster transformation) 
have less information.  Practically, it is difficult to find appropriate 
procedure of the inverse RG. The advantage of using the machine-learning 
is to find a rule to connect the renormalized configuration and 
the original configuration by searching for large amounts of datasets.  
This is the idea of the machine-learning. 
A single realization of $32 \times 32$ system in the present study 
does not have the information of larger systems. 
With the help of SR technique of machine-learning, we can obtain 
the information of larger systems which include the scaling properties. 

As a preliminary study, we are also considering the block-cluster 
transformation for the three-dimensional (3D) systems.  In the case 
of the simple cubic lattice, we determine a label of the block 
cluster from the labels of $2 \times 2 \times 2$ sites using 
a majority rule for the labels. The block-cluster 
transformation works very well for the 3D Ising model. 
For the 3D Ising model, a trial to improve convergence using 
the modified block-spin transformation~\cite{Blote} 
in the Monte Carlo RG calculation was 
reported recently~\cite{Ron2020}. 
In the Monte Carlo RG studies~\cite{Pawley,Baillie,Ron2020}, 
the correlations between different blocking levels $(m), (n)$, 
$\l s_{\gamma}^{(n)} s_{\beta}^{(m)} \r - 
 \l s_{\gamma}^{(n)} \r \l s_{\beta}^{(m)}\r$, 
are calculated. 
In the present formalism of block-cluster transformation, 
such correlations can be calculated.  It will be interesting to 
apply the block-cluster transformation, where first-order moments 
$\l s_{\beta} \r$ are automatically zero because of the improved 
estimator.

It is well-known that the wavelet transformation 
is a highly efficient representation of images by decomposing 
the image signal into high-frequency and low-frequency sub-bands. 
Guo {\it et al.}~\cite{Guo} proposed a deep wavelet SR method 
to recover missing details of low-resolution images.
Tomita~\cite{Tomita2018} performed the wavelet analysis of 
a configuration of FK clusters.  The SR study on spin models 
using wavelet transformation will be informative. 

For future studies, it will be interesting to treat continuous spin systems, 
such as the XY (clock) model, along the present scheme. 
The application of RG and inverse RG analyses to quantum systems \cite{Wu} 
is also challenging. 

\section*{Methods}

\subsection*{Detailed procedure of Block-cluster transformation}

\begin{figure}
\begin{center}
\includegraphics[width=7.8cm]{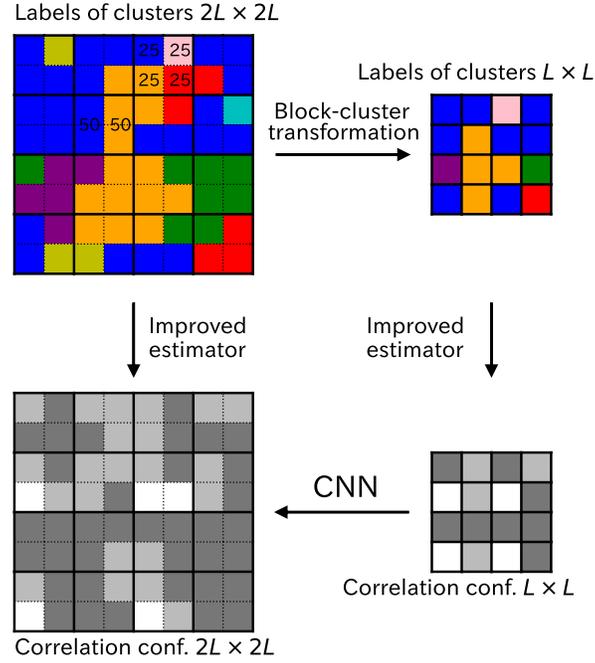}
\caption{
Schematic illustration of block-cluster transformation. 
We employ a majority rule to determine a label of the block cluster 
from the labels of $2 \times 2$ sites. 
The labels of clusters are represented by different colors. 
We also show the procedure of calculating improved correlation 
of the original configuration and the renormalized configuration 
together with the super-resolution process of CNN. 
The improved correlation configurations take a value of 
1 (white), 1/2 (light gray), or 0 (dark gray).
}
\label{fig:block_cluster}
\end{center}
\end{figure}

In the cluster update of Monte Carlo simulation, the spins are 
classified by the FK clusters~\cite{KF,FK}, 
and we assign a label to each FK cluster. 
We note that the connection between the magnetization of the $q$-state 
Potts model and the percolation probability of the cluster model 
was discussed by Hu\cite{Hu1984a,Hu1984b}. 
We employ a majority rule to determine a label of the block cluster 
from the labels of $2 \times 2$ sites. 
The procedure of the block-cluster transformation is schematically 
illustrated in Fig.~\ref{fig:block_cluster}, 
where different colors are used to assign the labels of clusters, 
not the spins. Even if adjacent sites have the same spins, these 
sites may have different colors because of the FK cluster.
If the labels of two sites coincide, we take this label as the label 
of the block.  The choice of two sites is sixfold.  
If there is no pair to coincide, we choose one label 
from four sites with 25\% probability. 
When two sites have one label and the other two sites have 
another label, one may pick up one with 50\% probability. 
However, we can take the label of the pair, which was first picked up, 
deterministically; it is the same situation as the two-up two-down 
case of the Ising-model block-spin 
transformation~\cite{Efthymiou}.
It is noteworthy that 
the procedure of block-cluster transformation is the same 
for $q$-state models irrespective of $q$.
In the framework of the dual Monte Carlo 
algorithm~\cite{Kandel,Kawashima95,Kawashima04}, 
the block-cluster transformation is a transformation in the 
graph degrees of freedom, whereas the block-spin transformation 
is that in the spin degrees of freedom. 
It is instructive to compare the program code for the block-cluster 
transformation with that for the block-spin transformation. 
The essential part to select the label of the block cluster is
\begin{verbatim}
   if(cluster[blk[la]] == cluster[blk[la]+1])
      cluster_blk[la] = cluster[blk[la]]; ,
\end{verbatim}
which is in parallel to the block-spin transformation:
\begin{verbatim}
   if(spin[blk[la]] == spin[blk[la]+1])
      spin_blk[la] = spin[blk[la]]; .
\end{verbatim}
In the program code, \verb+cluster[]+ 
stands for the label of the original cluster, 
whereas \verb+cluster_blk[]+ stands for that of 
the block cluster. The command means that if the two labels in the block 
coincide, we choose this label as the label of the block cluster.

\subsection*{Detailed description of SR}

We describe the detailed procedure of the SR with a deep CNN.  
It is an extension of the method by 
Efthymiou {\it et al.}~\cite{Efthymiou}. 
The improved estimator of the correlation configuration is used 
instead of the spin configuration. 
Starting from $2L \times 2L$ 
spin system of improved correlation configuration, $\{ \xi_i \}$,
we produce $L \times L$ system of improved correlation configuration, 
$\{ g_i \}$, using the block-cluster transformation. 
We aim to reproduce the $2L \times 2L$ target improved 
correlation configuration using the SR mapping of 
a supervised learning approach. 
This procedure is also illustrated in Fig.~\ref{fig:block_cluster}. 
The improved correlation configurations take a value of 
1, 1/2, or 0 because we consider the average value of $x$- and 
$y$-directions. According to Ref.~\citen{Tomita2020}, they are 
represented in white, light gray or dark gray. 

We consider three layers for the SR CNN, patch extraction 
and representation, non-linear mapping, and 
reconstruction~\cite{Dong}. 
The first layer is an upsampling layer 
by copying each configuration to $2 \times 2$ 
block. This upsampling procedure is appropriate 
both for low temperature and high temperature. 
It is in contrast to the situation of using spin configuration 
\cite{Efthymiou}, 
where a simple upsampling procedure is insufficient 
for random configuration of high-temperature side. 
For the improved correlation configuration, it takes $+1$ 
when the correlation length is as large as half of the 
system size, $L/2$, and the sites with $+1$ correlation form 
a small cluster (see Fig.~1(f) of Ref.~\citen{Tomita2020}). 
For a convolution layer, as the second layer, 
the transformation $f(W * x + b)$ is applied 
to the input $x$, a $2 \times 2$ improved correlation 
configuration. Here, $W$ ($2 \times 2$) is the filter, 
$b$ being the bias vector, and $*$ being the convolution operation. 
To avoid truncating the image edge, 
we add the periodic boundary padding. 
For activation function $f$, we use a sigmoid function, which gives 
the probability of each site 
For a loss function, we use the cross-entropy loss function 
between $\{ \xi_i \}$ and $\{ p_i \}$ (continuous variable); 
\begin{equation}
  L(\{\xi_i\}, \{p_i \}) =  - \sum_{i=1}^{N} 
  \Big[ \xi_i \cdot \ln p_i + (1 - \xi_i) \cdot \ln (1 - p_i ) \Big],
\end{equation}
where $\cdot$ denotes the element-wise product between matrices.
As a library we use "BCEWithLogitsLoss", where BCE stands 
for binary cross entropy. 
Parameters ($\theta=(W,b)$) are tuned to minimize 
a loss function. We employ the Adam method \cite{Adam} as an optimizer. 
As the third layer, using the optimized parameters, 
we calculate each $\{ p_i \}$, 
and determine $+1$ or $0$ depending upon this probability. 
Repeating this process two times, we emulate a configuration as 
the sum of $x$- and $y$-directions.

Thus, we can reproduce $\{ \xi_i' \}$ of $2L \times 2L$ size. 
Because we deal with the correlation, we can treat the three-state Potts 
model;  the permutational symmetry is taken into account. 
When we consider the spin configuration, we cannot follow 
the present super-resolution procedure. We emphasize that 
the same procedure can be used for any $q$-state Potts model.

\section*{Acknowledgements}
This work was supported by a Grant-in-Aid for Scientific Research 
from the Japan Society for the Promotion of Science, Grant Number, JP19K03657.
It was also supported by a Research Fellowships of Japan Society 
for the Promotion of Science for Young Scientists, Grant Number 20J12472. 
K. S. is grateful to the A*STAR (Agency for Science, Technology 
and Research) Research Attachment Programme of Singapore for financial support.
We generated eps files of figures using python library of matplotlib. 

\section*{Author contributions}
K. S. and Y. O. designed the study and performed computer calculations. 
All authors analyzed the results, and approved the final manuscript.

\section*{Competing interests}
The authors declare no competing interests.

\end{document}